\documentclass{emulateapj}
\usepackage{amsmath}
\usepackage{amssymb}
\usepackage{amsthm}
\usepackage{array}
\usepackage{float}
\usepackage{graphicx}
\usepackage{subfigure}
\usepackage{color}

\newcommand{\myemail}{kzk15@psu.edu}

\slugcomment{}

\shorttitle{UV/optical/IR transients after BSG GRBs}
\shortauthors{Kashiyama et al.}

\begin{document}

\title{Luminous supernova-like UV/optical/infrared transients associated with \\ ultra-long gamma-ray bursts from metal-poor blue supergiants} 

\author{Kazumi Kashiyama\altaffilmark{1}, Daisuke Nakauchi\altaffilmark{2}, Yudai Suwa\altaffilmark{3}, Hidenobu Yajima\altaffilmark{1}, and Takashi Nakamura\altaffilmark{2}}
\email{\myemail}

\altaffiltext{1}{Department of Astronomy \& Astrophysics; Department of Physics; Center for Particle \& Gravitational Astrophysics; Pennsylvania State University, University Park, PA 16802}
\altaffiltext{2}{Department of Physics, Kyoto University, Oiwake-cho, Kitashirakawa, Sakyo-ku, Kyoto 606-8502, Japan}
\altaffiltext{3}{Yukawa Institute for Theoretical Physics, Kyoto University, Oiwake-cho, Kitashirakawa, Sakyo-ku, Kyoto 606-8502, Japan}

\begin{abstract}   
Metal-poor massive stars may typically end up their lives as blue supergiants (BSGs).
Gamma-ray bursts (GRBs) from such progenitors could have ultra-long duration of relativistic jets.  
For example Population III (Pop III) GRBs  at $z \sim10\mbox{-}20$ might be observable as X-ray rich events with a typical duration  of $T_{90}\sim 10^4(1+z) \ {\rm sec}$.  
Recent GRB111209A at $z = 0.677$ has an ultra-long duration of $T_{90}\sim 2.5\times 10^4 \ {\rm sec}$ 
so that it have been suggested that the progenitor might be a metal-poor BSGs in the local universe. 
Here, we suggest luminous UV/optical/infrared emissions associated with such a new class of GRB from metal poor BSGs.   
Before the jet head breaks out the progenitor envelope, the energy injected by the jet is stored in a hot-plasma cocoon, 
which finally emerges and expands as a baryon-loaded fireball.  
We show that  the photospheric emissions from the cocoon fireball could be intrinsically very bright ($L_{\rm peak} \sim 10^{42\mbox{-}44} \ \rm erg/sec$) 
in UV/optical bands ($\varepsilon_{\rm peak} \sim 10 \ \rm eV$) with a typical duration of $\sim 100 \ \rm days$ in the rest frame. 
Such cocoon emissions from Pop III GRB might be detectable in infrared bands at $\sim$ years after Pop III GRBs at up to $z \sim 15$ by up-coming facilities like JWST. 
We also suggest that GRB111209A might have been rebrightening in UV/optical bands up to an AB magnitude of $\lesssim 26$.  
The cocoon emissions from local metal-poor BSGs might have been already observed as luminous supernovae without GRB since they can be seen from the off-axis direction of the jet. 
\end{abstract}


\section{Introduction}
Direct signals from Population III (Pop III) stars are needed to confirm their existence 
as well as to know the earliest history of star formation, galaxy evolution, and cosmic reionization at $z \gtrsim 10$.  
However, to observe Pop III stars in their stellar phase is extremely difficult due to the distance.  
Therefore the core-collapse phase may be more suitable to observe Pop III stars.  
Especially possible supernovae and GRBs by Pop III stars are good candidates.

Now Pop III stars are considered to be formed in a dark-matter halo of $\sim 10^6 M_{\odot}$ at $z \gtrsim 20$.  
The molecular hydrogen can cool the gas to $\sim 200 {\rm K}$, which determines the onset of the Jeans instability to form gas clumps of $\sim 10^3 M_{\odot}$ 
\citep[e.g.,][]{Bromm_et_al_1999,Nakamura_Umemura_2002}.  
If the most mass is accreted to form the star, the Pop III star at the zero-age main sequence (ZAMS) would be as massive as $M_{\rm ZAMS} \sim10^{2}$--$10^3 M_{\odot}$.  
If so, there have been discussed that Pop III stars with ZAMS mass in a range of $140\mbox{-}260 M_{\odot}$ would result in pair-instability supernovae (PISNe) \citep{Woosley_et_al_2002}.
Detailed numerical simulations have shown that the signals can be detected in infrared bands up to $z \sim 30$ using JWST \citep[e.g.][]{Whalen_et_al_2012a}
\footnote{Even in the case of the lighter mass range $\lesssim 40 M_{\odot}$, future facilities may catch the signal of type II SN up to $z \sim 15$ \citep{Whalen_et_al_2012b}.}. 
As for even more massive stars, several core-collapse simulations were performed, which would produce black holes with $\sim O(100) M_{\odot}$ 
\citep[e.g.][]{Fryer_et_al_2001, Suwa_et_al_2007a, Suwa_et_al_2007b}.

However, such large ZAMS mass Pop III stars seem to be formed only for spherically symmetric systems so that ZAMS mass of Pop III stars may change when we take effects of rotation into account.
Some cosmological simulations indicate that the rotation of these gas clumps can naturally split them into sub clumps of $\lesssim 100 M_{\odot}$ 
\citep{Turk_et_al_2009,Stacy_et_al_2010,Clark_et_al_2011}.
Even if the seed gas clump is more massive, the protostellar UV radiation evaporates a significant fraction of the mass 
so that the Pop III stars are as small as $M_{\rm ZAMS} \lesssim (30-90) M_{\odot}$ \citep{McKee_Tan_2008,Hosokawa_et_al_2011}.  
Population III.2 stars, which may be more abundant, can also have smaller ZAMS mass \citep[e.g.,][]{Yoshida_et_al_2007}. 
Massive Pop III stars typically end up their lives as blue supergiants (BSGs) with massive hydrogen envelope due to the low opacity in the envelope \citep{Woosley_et_al_2002}.

For such BSG Pop III stars, several authors have discussed the possibility of gamma-ray bursts (GRBs) 
\citep{Meszaros_Rees_2010, Komissarov_Barkov_2010,Suwa_Ioka_2011,Nagakura_et_al_2011,Woosley_Heger_2012}, 
which we like to discuss from another point.  
One of the most promising scenarios of long GRBs is {\it collapsar} scenario, in which relativistic jets burrowing through the stellar envelope 
are indispensable to bring relativistically moving elements into optically thin region without dissipation. 
For massive progenitors with enough angular momentum, a stellar mass black hole (BH) with an accretion disk would be first formed in the core-collapse phase 
\citep{Woosley_1993,Narayan_et_al_2001}, 
and relativistic jets may be launched via the Blandford-Znajeck (BZ) like process \citep{Blandford_Znajek_1977,McKinney_Gammie_2004,Meszaros_Rees_2010}
\footnote{Another alternative of the jet production mechanism is neutrino process \citep{MacFadyen_Woosley_1999, Popham_et_al_1999, Zalamea_Beloborodov_2011, Suwa_2012}, 
in which copious amount of neutrinos and anti-neutrinos emitted from the hyperaccreting disk annihilate in the vicinity of the axis and produce a fireball consisting of hot electron-positron pairs.}.  
If the jet can break out the stellar envelope before the central engine ceases, it may produce prompt gamma rays by some dissipation processes near and beyond the photospheric radius 
\citep[see e.g.,][]{Meszaros_2012}.
Also, one can expect radio to infrared afterglows via synchrotron emissions from the non-thermal electrons accelerated at the shocks in the relativistic shell going through the circumstellar medium
\citep{Ioka_Meszaros_2005, Inoue_et_al_2007, Toma_et_al_2011, Nakauchi_et_al_2012}.

In terms of whether the jet can break out the envelope or not, the anticipated large pre-collapse mass of Pop III stars was thought to be problematic.  
Indeed, \cite{Matzner_2003} claimed that the progenitors of the observed long GRB would not be giant stars like red supergiants (RSGs) but compact stars like CO Wolf-Rayet (WR) stars.  
Nevertheless several authors have claimed that Pop III stars can produce GRBs irrespective of the pre-collapse mass 
\citep{Meszaros_Rees_2010, Suwa_Ioka_2011,Nagakura_et_al_2011,Woosley_Heger_2012}.  
This is essentially because they would be BSGs in the pre-collapse phase \citep{Meszaros_Rees_2001}. 
The large mass gives a larger intrinsic energy budget, and the relative compactness for the large mass helps the jet to reach the stellar edge before the engine ceases.  
As an interesting outcome, the duration of the prompt emission of Pop III or BSG GRBs would be much longer $\gtrsim 10^4(1+z) \ {\rm sec}$ compared with the observed value 
\citep{Suwa_Ioka_2011,Nakauchi_et_al_2012}.
\cite{Nakauchi_et_al_2012} argued the detectability of such ultra-long GRBs from BSGs and found that 
if the peak spectral energy and isotropic equivalent energy ($E_p-E_{\rm iso}$) correlation holds the observed $E_p^{\rm obs}\sim 100 \ {\rm keV}$ with $E_{\rm iso}\sim 10^{54} \ {\rm erg}$ 
and the duration $T_{90}=6\times 10^4 \ {\rm sec}$ for Pop III BSG with mass $40 M_\odot$ at $z=9$ which can be detected by an ${\it EXIST}$-type future instrument.  
These values are similar to the recent observation of ultra-long GRB111209A with $E_{\rm iso}=5.82\times 10^{53} \ {\rm erg}$, $E_p^{\rm obs}= 310 \ {\rm keV}$ 
and the duration of $\sim 2.5\times 10^4 \ {\rm sec}$ \citep{Gendre_et_al_2012}.  
Given that the observed redshift is relatively small, $z=0.677$, the progenitor of this GRB might be a metal-poor BSG in the local universe.

Here, we propose a new possible photon emission associated with such ultra-long GRBs from Pop III stars, or more general BSG progenitors.
We consider a cocoon i.e., a hot-plasma sheath of the jet, which is a necessary ingredient of the above collapsar model of GRB.  
Before the jet breaks out the progenitor, the energy is stored in the cocoon \citep[e.g., ][]{Lazzati_Begelman_2005,Bromberg_et_al_2011}, 
which can finally emerges and expands as a baryon-loaded fireball.  
We evaluate photospheric emissions from such cocoon fireballs for various progenitors and discuss the detectability. 
We show that the emission is sensitive to the progenitor, so the cocoon fireball photospheric emission (hereafter CFPE) can be a diagnostic of Pop III and BSG GRBs.

This paper is organized as follows.  In Sec.\ref{sec:pro}, we introduce GRB progenitor models which we use in the following calculations. 
In Sec.\ref{sec:cocoon_form_evo}, we model jet-cocoon formation and evolution before and after breaking out each progenitor.
In Sec.\ref{sec:cocoon_emi}, we model CFPE, and evaluate the detectability.  
In Sec.\ref{sec:sum_dis}, we summarize our calculations.
 
\section{Progenitor model}\label{sec:pro}

\begin{table*}
\begin{center}
\caption{Cocoon parameters at the jet breakout and the photon emissions at the photospheric radius. }
\label{table_1}
{\renewcommand\arraystretch{1.6}
\begin{tabular}{ccc|c|ccccc|ccc}
\hline 
  \shortstack{Progenitor \\ model} & \shortstack{$M_\ast$ \\ $[M_{\odot}]$} &  \shortstack{$R_{\ast}$ \\ $\rm [cm]$} & $(\alpha,\eta_{\rm j},\theta_{\rm j})$ & \shortstack{$E_{\rm c, bo}$ \\ $\rm [erg]$ } & \shortstack{$V_{\rm c, bo}$ \\ $[{\rm cm^3}]$} & \shortstack{$M_{\rm c, bo}$ \\ $[M_{\odot}]$} & $\tau_{\rm c, bo}$ & $\eta_{\rm c}$ & \shortstack{$L_{\rm peak}$ \\ $\rm [erg/sec]$ } & \shortstack{$\varepsilon_{\rm peak}$ \\ $\rm [eV]$} & \shortstack{$t_{\rm peak}$ \\ $\rm [day]$}  \\ \hline
z40BSG$^{\dagger}$ & $40$ & $1.5 \times 10^{12}$ & & $1.8 \times 10^{52}$ & $3.2 \times 10^{35}$ & $0.92$ & $3.4 \times 10^9$ & $0.011$ & $2.4 \times 10^{42}$ & $13$ & $87$ \\ 
z70BSG$^{\ddagger}$ & $70$ & $1.4 \times 10^{13}$ & & $3.9 \times 10^{52} $ & $1.5 \times 10^{38}$ & $0.88$ & $6.7 \times 10^7$ & $0.025$ & $1.6 \times 10^{43}$ & $11$ & $94$  \\
z915BSG$^{\diamondsuit}$ & $915$ & $8.8 \times 10^{12}$ & & $6.2 \times 10^{53}$ & $5.0 \times 10^{37}$ & $16$ & $2.2 \times 10^9$ & $0.022$ & $2.1 \times 10^{43}$ & $9.5$ & $350$  \\  
{\it l}40BSG$^{\dagger}$ & $40$ & $4.4 \times 10^{12}$ &  fiducial & $1.8 \times 10^{52}$ & $1.0 \times 10^{37}$ & $1.1$ & $3.9 \times 10^8$ & $8.8 \times 10^{-3}$ & $4.8 \times 10^{42}$ & $12$ & $91$  \\  
{\it l}75BSG$^{\dagger}$ & $75$ & $8.6 \times 10^{12}$ & & $4.5 \times 10^{52}$ & $7.5 \times 10^{37}$ & $2.1$ & $1.9 \times 10^8$ & $0.012$ & $1.2 \times 10^{43}$ & $11$ & $120$  \\ 
s40WR$^{\dagger}$ & $8.7$ & $2.3 \times 10^{10}$ & & $2.3 \times 10^{51}$ & $6.5 \times 10^{29}$ & $0.11$ & $3.1 \times 10^{12}$ & $0.012$  & $6.8 \times 10^{40}$ & $22$ & $40$ \\ 
s75WR$^{\dagger}$ & $6.3$ & $3.7 \times 10^{10}$ & & $1.1 \times 10^{51}$ & $2.5 \times 10^{30}$ & $0.075$ & $8.9 \times 10^{11}$ & $8.0 \times 10^{-3}$  & $5.5 \times 10^{40}$ & $21$ & $37$ \\  \hline
{\it l}75BSG &  &  & $\eta_{\rm j} \times 10$ & $3.8 \times 10^{53}$ & $7.3 \times 10^{37}$ & $2.0$ & $1.9 \times 10^8$ & $0.10$ & $1.2 \times 10^{44}$ & $13$ & $68$  \\ 
{\it l}75BSG &  &  & $\eta_{\rm j} \times 1/10$ & $6.0 \times 10^{51}$ & $1.2 \times 10^{38}$ & $3.4$ & $1.9 \times 10^8$ & $9.7 \times 10^{-4}$ & $1.2 \times 10^{42}$ & $9.1$ & $220$  \\ 
{\it l}75BSG &  &  & $\theta_{\rm j} \times 2$ & $5.3 \times 10^{52}$ & $1.9 \times 10^{38}$ & $5.3$ & $1.9 \times 10^8$ & $5.6 \times 10^{-3}$ & $1.3 \times 10^{43}$ & $10$ & $140$  \\  
{\it l}75BSG &  &  & $\theta_{\rm j} \times 1/2$ & $4.0 \times 10^{52}$ & $3.4 \times 10^{37}$ & $0.96$ & $1.9 \times 10^8$ & $0.023$ & $1.1 \times 10^{43}$ & $11$ & $98$  \\  \hline
\end{tabular}
}
\\
\end{center}
The first column shows model names which include a letter representing metallicity (z-zero, l-low, s-solar), the ZAMS mass, and the pre-collapse state (BSG or WR). 
The second and third columns show the mass and the radius of the progenitors, respectively.  
The 4th column shows the parameter of our theoretical model. 
Fiducial parameter means $(\alpha, \eta_{\rm j}, \theta) = (1,6.2 \times 10^{-4}, 0.1)$. 
The 5th to 12th columns show the energy, the volume, the mass and the optical depth of the cocoon at the jet breakout, the baryon load parameter,
the peak luminosity, the peak energy and the peak time of cocoon-fireball photospheric emission (CFPE), respectively.
For details, see the text. \\
The reference: ${}^{\dagger}$\cite{Woosley_et_al_2002}, ${}^{\ddagger}$\cite{Heger_Woosley_2010}, ${}^{\diamondsuit}$\cite{Ohkubo_et_al_2009}
\begin{center}
\end{center}
\end{table*}

We consider massive stars ($M_{\rm ZAMS} \geq 40 M_{\odot}$) with zero- ($Z = 0$), low- ($Z = 10^{-4} Z_{\odot}$), and also solar metallicity ($Z=Z_{\odot}$). 
We use pre-collapse stellar models given by \cite{Woosley_et_al_2002}, \cite{Heger_Woosley_2010} and \cite{Ohkubo_et_al_2009}, which is listed in Table 1.  
The first column shows model names which include a letter representing metallicity (z-zero, l-low, s-solar), the ZAMS mass, and the pre-collapse state (BSG or WR).  
The mass ($M_\ast$) and the radius ($R_\ast$) in the pre-collapse phase are shown in the second and the third columns, respectively. 
One can see that only the progenitors with solar metallicity lose the dominant masses before collapsing to become WR stars. 
This is due to the strong stellar wind blowing out the hydrogen envelope, which may not be the case for zero- or low-metal stars due to the low opacity in the envelope.  
Such mass losses also make the WR progenitors much more compact than the BSG progenitors.

\section{Cocoon formation and evolution} \label{sec:cocoon_form_evo}

\subsection{Before jet-cocoon breakouts}
In the case of massive progenitors with the pre-collapse mass $\gtrsim 40 M_\odot$, a stellar mass BH would be first formed just after the collapse is triggered \citep{Heger_et_al_2003}.  
We assume that the initial BH mass is $3 M_{\odot}$, and evaluate the time-dependent mass accretion onto the BH as
\begin{equation}\label{eq:M_dot}
\dot{M}(t) = \alpha \frac{dM_{r}}{dt_{\rm ff}} \Big|_{t = t_{\rm ff}(r)}.
\end{equation}
Here $M_{r} = \int_0^{r} 4 \pi r'{}^2 \rho_\ast(r') dr'$ is the mass coordinate for $M_r > 3 M_{\odot}$ with $\rho_\ast(r)$ being the density of the steller envelope, 
$t_{\rm ff} = (r{}^3/GM_{r})^{1/2}$ is the free-fall time of a mass shell $dM_{r}$ with $G$ being the gravitational constant, and $\alpha \lesssim 1$ represents the deviation from the free-fall accretion
\footnote{In \cite{Suwa_Ioka_2011} and \cite{Nakauchi_et_al_2012}, the suppression factor of accretion rate, $\alpha$, was absorbed in the definition of the jet efficiency $\eta_{\rm j}$; 
their $\eta_{\rm j}$ corresponds to $\alpha\eta_{\rm j}$ in this paper.}, 
which may be time-dependent \citep[see e.g.,][]{Kumer_et_al_2008}.  
We use $t$ as a time coordinate from the start of the accretion in the rest frame of the central engine.

We assume that the luminosity of the relativistic jet is proportional to the mass accretion rate into the central BH;
\begin{equation}\label{eq:L_iso}
L_{\rm j}(t) = \eta_{\rm j} \dot{M}(t) c^2,
\end{equation}
where $\eta_{\rm j}$ is the efficiency factor and $c$ is the speed of light. 
Eq. \eqref{eq:L_iso} can be realized in the BZ like process \citep[e.g.,][]{Komissarov_Barkov_2010,Kawanaka_et_al_2012}.  
Note that $\eta_{\rm j}$ may be also time-dependent.

When the jet collides with the stellar envelope, two shocks are formed; a forward shock propagating in the stellar envelope, and a reverse shock in the jet.  
We call the region sandwiched by the two shocks as jet head.  
We consider jets with a Lorentz factor $\Gamma_{\rm j} \gtrsim 10$ ($\beta_{\rm j} \simeq 1$).  
In the thin-shell limit \citep{Sari_Piran_1995,Meszaros_Waxman_2001}, which would be valid in our cases
\footnote{The sufficient condition is $\hat{L}(t) \ll \Gamma_{\rm j}{}^4 $.}, 
the two shocks approximately proceeds with a single Lorentz factor $\Gamma_{\rm h}$.  
Then, the pressure balance at the jet head, 
$(L_{\rm j}/\pi \theta_{\rm j}^2 \Gamma_{\rm j}{}^2 r_{\rm h}{}^2 c) \times (\Gamma_{\rm j} \Gamma_{\rm h})^2 (1-\beta_{\rm h})^2 = \rho_{\ast} c^2 \Gamma_{\rm h}{}^2 \beta_{\rm h}^2$ 
yields the velocity of the jet head as
\begin{equation}
\beta_{\rm h}(t) = \frac{1}{1 + \hat{L}(t)^{-1/2}},
\end{equation}
where 
\begin{equation}
\hat{L}(t) \equiv \frac{L_{\rm j}(t - r_{\rm h}/c)}{\pi \theta_{\rm j}^2 r_{\rm h}{}^2 \rho_\ast(r_{\rm h}) c}, 
\end{equation}
and $\theta_{\rm j}$ is the half opening angle of the jet.  
Here, the subscript ``$\rm j$'' and ``$\rm h$'' refer the jet and the jet head, respectively.  
The position of the jet head is obtained from $r_{\rm h} = \int_0^t c\beta_{\rm h} dt$.

As far as the jet head is non-relativistic, $\beta_{\rm h} \lesssim 0.3$, 
a dominant fraction of shocked plasma at the jet head will spread out sideways to form a cocoon \citep[e.g.,][]{Bromberg_et_al_2011}.  
The energy stored in the cocoon is evaluated as
\begin{equation}
E_{\rm c}(t) = \int^t L_{\rm j}(t' -r_{\rm h}(t')/c) d t'.
\end{equation}
Hereafter we simply assume the shape of the cocoon as circular cone with the height of $\approx r_{\rm h}(t)$, 
and with the base radius of $r_{\rm c}(t)$, where the subscript ``$\rm c$'' refers to the cocoon.
Then the volume of the cocoon can be evaluated as
\begin{equation}
V_{\rm c}(t) = \frac{\pi}{3} r_{\rm c}(t){}^2 r_{\rm h}(t).
\end{equation} 
The pressure balance at the cocoon-progenitor interface, $E_{\rm c}/3 V_{\rm c} = \rho_\ast \beta_{\rm c}{}^2 c^2$, yields the transverse expansion velocity of the cocoon as
\begin{equation}
\beta_{\rm c}(t) = \left( \frac{E_{\rm c}(t)}{3 \rho_\ast(r_{\rm h}) c^2 V_{\rm c}(t)}\right)^{1/2}.
\end{equation}
The transverse size of the cocoon is given by $r_{\rm c} = \int_0^t c\beta_{\rm c} dt$.  
Finally, one can evaluate the baryon mass loaded in the cocoon as
\begin{equation}\label{eq:Mc}
M_{\rm c}(t) = \frac{r_{\rm c}(t){}^2}{4r_{\rm h}(t){}^2} \int^{r_{\rm h}(t)} 4 \pi r^2 \rho_*(r)dr,
\end{equation}
In principle, the cocoon is divided into an inner cocoon consisting of the shocked-jet matter and an outer cocoon of the shocked stellar matter.  
The two region can be separated by a contact discontinuity \citep[e.g.,][]{Bromberg_et_al_2011}.  
In Eq. (\ref{eq:Mc}), we assume that the contact surface becomes unstable due to e.g., a Kelvin-Helmholtz type instability, and the outer and the inner cocoon are fully mixed.
We discuss the cases where such cocoon mixing is insufficient in Sec.\ref{sec:sum_dis}. 

If the jet head arrives at the stellar surface before the central engine ceases, that is $t_{\rm bo} < t_{\rm ff}(R_{\ast})/\alpha$, 
the jet succeeds to break out the stellar envelope, where $t_{\rm bo}$ is defined by $r_{\rm h}(t_{\rm bo})=R_{\ast}$.  
Hereafter the subscript ``$\rm bo$'' stands for {\it breakout}.  Note that the jet can successfully break out the stellar envelope for all the progenitor models shown in Table 1.

In our prescription, the three parameters $\alpha$, $\eta_{\rm j}$, and $\theta_{\rm j}$ determine whether the jet breaks out the envelope or not as well as parameters of the cocoon.  
In general, both $\alpha$ and $\eta_{\rm j}$ depend on time, due to the details of the accretion disk formation and the magnetic field structure around the BH.
The jet-opening angle $\theta_{\rm j}$ also depends on time due to the compression by the cocoon pressure \citep[e.g.,][]{Lazzati_Begelman_2005,Bromberg_et_al_2011}.  
In the following calculation, we use $\alpha = 1$, $\eta_{\rm j} = 6.2 \times 10^{-4}$, and $\theta_{\rm j} = 0.1$ as fiducial values.
By applying these to WR progenitors, it is shown that  the observed characteristics of canonical GRBs can be reproduced \citep{Suwa_Ioka_2011}.

\begin{figure}
\includegraphics[width=90mm]{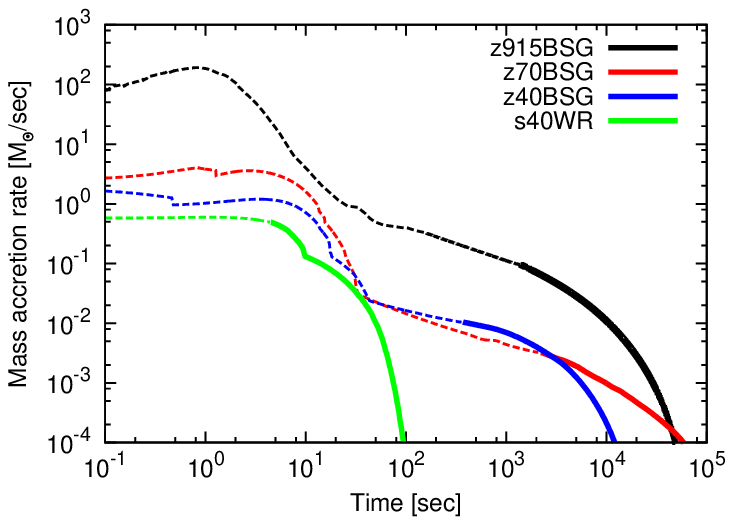}
\caption{The mass accretion rate in the central-engine rest frame of the progenitor models in Table. 1. 
In each line, the dotted and the solid region corresponds to the time before and after the jet breaks out the progenitor, respectively.
We fix $\alpha = 1$ (see Eq.(\ref{eq:L_iso})). 
}
\label{mdot}
\end{figure}

Fig. \ref{mdot} shows the mass-accretion rate in the central-engine rest frame of each progenitor model in Table. 1. 
The dotted and the solid regions correspond to the time before and after the jets break out the progenitors, respectively. 
One can see that metal-poor progenitors share some characteristics. 
The accretion rate is initially as high as $\sim 1\mbox{-}100 \ M_{\odot}/\rm sec$, 
which suddenly decreases down by $\sim 10^{-2}$ at $t \sim 10 \ \rm sec$. 
This transition corresponds to the end of the accretion of the He core. 
After that, the accretion rate shallowly decreases as approximately $\propto t^{-0.73}$ where the hydrogen envelope is accreted \citep{Suwa_Ioka_2011}. 
The jet typically breaks out the envelope at $t \sim 10^3 \ \rm sec$.  
Even at this time, the accretion rate is still $\gtrsim 10^{-3} \ M_{\odot}/\rm sec$ 
which may maintain the relativistic-jet activity  \citep{Chen_Beloborodov_2007,Kawanaka_et_al_2012}.
The accretion essentially ends at $t \gtrsim 10^{4} \ \rm sec$, which is consistent with the observed duration of the ultra-long GRB. 
On the other hand, in the case of the WR progenitor, the jet breakout occurs at $t \lesssim 10 \ \rm sec$, and the massive accretion ends at $t \sim 10^2 \ \rm sec$. 
This is consistent with the observed duration of typical long GRB. 

In Table \ref{table_1}, the internal energy ($E_{\rm c, bo} \equiv E_{\rm c}(t_{\rm bo})$), volume ($V_{\rm c, bo}\equiv V_{\rm c}(t_{\rm bo})$), 
and the baryon mass of the cocoon ($M_{\rm c, bo} \equiv V_{\rm c}(t_{\rm bo})$) at the jet-cocoon breakout for various progenitors are shown.  
We also show the so called baryon-load parameter of the cocoon,
\begin{equation}
\eta_{\rm c} \equiv \frac{E_{\rm c,bo}}{M_{\rm c,bo} c^2},
\end{equation}
and the optical depth of the cocoon at the breakout, 
\begin{equation}
\tau_{\rm c, bo} = \frac{\sigma_{\rm T} M_{\rm c,bo} R_{\ast}}{m_{\rm p}V_{\rm c,bo}}.
\end{equation}
Here $\sigma_\mathrm{T} = 6.7 \times 10^{-25} \mathrm{cm^2}$ is the Thomson cross section and $m_\mathrm{p}$ is the proton mass.  
We should note that the cocoons are all {\it non-relativistic} 
\footnote{In e.g., \cite{Peer_et_al_2006}, \cite{Toma_et_al_2009}, and \cite{Starling_et_al_2012}, the effects of {\it relativistic} cocoon fireballs have been discussed 
to explain the characteristics of observed GRBs.  Cocoon fireballs can be relativistic if the mixing between inner and outer cocoon is suppressed.}, 
i.e., $\eta_{\rm c} < 1$, and also the cocoon fireballs are highly optically thick at the breakout, i.e., $\tau_{\rm c,bo} \gg 1$.

\subsection{After jet-cocoon breakouts}
As shown below, the evolution of the cocoon after the jet-cocoon breaking out the progenitor can be characterized by $(R_\ast, E_{\rm c,bo}, M_{\rm c,bo}, V_{\rm c,bo})$.  
We do not consider an additional energy injection to the cocoon from the jet after the breakout for simplicity.

The temperature of the cocoon fireball at the breakout can be determined from $E_{\rm c,bo} = a T_{\rm c,bo}{}^4 V_{\rm c,bo} + M_{\rm c, bo} k_{\rm B} T_{\rm c,bo} /m_{\rm p} $.  
In the case of cocoon fireball, the first term in the right-hand side is always dominant i.e., the fireball is radiation dominated, where the temperature evolution is described as 
$T_{\rm c} = T_{\rm c,bo} (V_{\rm c}/V_{\rm c,bo})^{-1/3}$ with
\begin{equation}
T_{\rm c,bo} = \left( \frac{E_{\rm c,bo}}{a V_{\rm c,bo}} \right)^{1/4}.
\end{equation}
Here $a = 7.6 \times 10^{-15} \mathrm{erg/cm^3/K^4}$ and $k_\mathrm{B} = 1.4 \times 10^{-16} \mathrm{erg/K}$ is the radiation constant 
and the Boltzmann constant, respectively.  Then, the radiation energy evolves as $aT_{\rm c}{}^4 V_{\rm c} \propto (V_{\rm c}/V_{\rm c,bo})^{-1/3}$, 
and the gas energy evolves as $(M_{\rm c,bo}/m_{\rm p})k_\mathrm{B} T_{\rm c} \propto (V_{\rm c}/V_{\rm c,bo})^{-1/3}$.
Since the scaling is the same, if the radiation dominates initially, the situation will remain as far as the cocoon is optically thick.

At first, the fireball would expand in an almost spherically symmetric manner, and the volume and the optical depth evolve as 
$V_{\rm c} = V_{\rm c, bo} (R/R_{\ast})^3$, $\tau_{\rm c}= \sigma_{\rm T} M_{\rm c,bo} R /m_{\rm p}V_{\rm c} = \tau_{\rm c, bo} (R/R_{\ast})^{-2}$, where $R$ is the radius of the fireball.  
As long as the cocoon is optically thick, the kinetic energy $K_{\rm c}$ increases with the volume as 
$dK_{\rm c} \approx p_{\rm c} dV_{\rm c} \approx aT_{\rm c}{}^4 dV_{\rm c}/3 = a T_{\rm c,bo}{}^4 V_{\rm c,bo} (R/R_{\ast})^{-2} d(R/R_{\ast})$, 
which gives $K_{\rm c}(R) = E_{\rm c,bo} (1-(R/R_{\ast})^{-1})$.  
This means that once the cocoon fireball expands twice as large as the initial size,
\begin{equation}
R_{\rm sat} \approx 2 R_{\ast}, 
\end{equation}
a considerable fraction of the internal energy will be delivered to the kinetic energy.  
The saturation velocity of the cocoon fireball is $v_\mathrm{sat} \approx ( 2 E_{\rm c,bo}/M_{\rm c,bo})^{1/2} = \sqrt{2\eta_{\rm c}} c $.  
The fluctuation of the velocity within the fireball $\delta v \approx (k_\mathrm{B}T_{\rm c}/m_\mathrm{p})^{1/2} < (k_\mathrm{B}T_{\rm c, bo}/2m_\mathrm{p})^{1/2}$ 
is always much smaller than $v_\mathrm{sat}$; $\delta v/v_\mathrm{sat} < (M_{\rm c,bo} k_\mathrm{B} T_{\rm c,bo}/m_\mathrm{p} E_{\rm c,bo})^{1/2} \ll 1$. 
Thus, the cocoon fireball can be approximated as a shell with a width $\Delta_R \approx R_\mathrm{sat}$ beyond the saturation radius,
where the volume and the temperature evolve as $V_{\rm c} \propto (R/R_\mathrm{sat})^2$, $T_{\rm c} \propto (R/R_\mathrm{sat})^{-2/3}$, respectively.  
The optical depth still evolves as $\tau_{\rm c}= \tau_{\rm c, bo} (R/R_{\ast})^{-2}$.  
Finally, the diffusion velocity of photons within the shell becomes equal to the coasting velocity of the shell , that is, $c/\tau_{\rm c}\approx v_\mathrm{sat}$. 
This occurs at the photospheric radius given by
\begin{equation}
R_\mathrm{ph} \approx 1.2 R_{\ast} \times (\tau_{\rm c, bo}{}^2 \eta_{\rm c})^{1/4}.
\end{equation}
In summary, cocoon fireballs evolve as
\begin{eqnarray}
V_{\rm c} = V_{\rm c,bo} 
\left\{\begin{array}{ll}
\left(\frac{R}{R_{\ast}}\right)^{-3} & \ (R_\ast < R < R_{\rm sat}) \\
\left(\frac{R_{\rm sat}}{R_{\ast}}\right)^{-3} \left(\frac{R}{R_{\rm sat}}\right)^{-2} & \ (R_{\rm sat} < R < R_{\rm ph} ), \\
\end{array} \right.
\end{eqnarray}
\begin{eqnarray}
T_{\rm c} = T_{\rm c,bo} 
\left\{\begin{array}{ll}
\left(\frac{R}{R_{\ast}}\right)^{-1} & \ (R_\ast < R < R_{\rm sat}) \\
\left(\frac{R_{\rm sat}}{R_{\ast}}\right)^{-1} \left(\frac{R}{R_{\rm sat}}\right)^{-2/3} & \ (R_{\rm sat} < R < R_{\rm ph} ), \\
\end{array} \right.
\end{eqnarray}
\begin{equation}
\tau_{\rm c}= \tau_{\rm c, bo} \left(\frac{R}{R_\ast}\right)^{-2}.
\end{equation}

\section{cocoon-fireball photospheric emissions}\label{sec:cocoon_emi}
In this section, we evaluate the photon emissions from the cocoon fireball.  
Here we only consider thermal photons and neglect other photon injection process, e.g. electron-synchrotron, bremsstrahlung emissions, which would give minor contributions in our cases.  
Also we neglect gamma rays as a decay product of unstable nuclei like ${}^{56}{\rm Ni}$.  
In fact \cite{Tominaga_et_al_2007} showed that the abundance of ${}^{56}{\rm Ni}$ synthesized by relativistic jets would be small.  
Moreover we neglect the effect of absorptions by nuclei included inside the fireball for simplicity.

In the coasting phase, thermal photons within a width of $\approx \Delta_R c/\tau_{\rm c}v_{\rm sat}$ can escape the shell, 
since they have a larger diffusion velocity than the coasting velocity of the shell. 
For a fixed radius $R$, the radiation first comes from near the line of sight and later from the limb so that the photons from the fixed $R$ are observed with a duration $\approx R/v_{\rm sat}$.  
Then, the mean bolometric luminosity can be evaluated as 
$L_{\rm b} \approx a T_{\rm c}{}^4 \times 4 \pi R^2 (\Delta_{R} c/\tau_{\rm c}v_{\rm sat}) \times (R/v_{\rm sat})^{-1} \propto (R/R_{\rm sat})^{1/3} \propto t^{1/3}$, 
which becomes the maximum at the photospheric radius, $R \approx R_{\rm ph}$.  
The time scale of the emission is estimated to be $t_{\rm peak} \approx R_{\rm ph}/v_{\rm sat}$ as
\begin{equation}\label{eq:tpeak}
t_{\rm peak} = 0.84 (R_{\ast}/c) \times (\tau_{\rm c, bo}{}^2/\eta_{\rm c})^{1/4}.  
\end{equation}
The peak bolometric luminosity is described as $L_{\rm peak} \approx E_{\rm c,bo} \times (R_{\rm sat}/R_{\ast})^{-1} (R_{\rm ph}/R_{\rm sat})^{-2/3} \times t_{\rm peak}{}^{-1}$, or
\begin{equation}\label{eq:Lpeak}
L_{\rm peak} = 0.84 (E_{\rm c,bo} c /R_{\ast}) \times \tau_{\rm c, bo}^{-5/6} \eta_{\rm c}^{1/12}, 
\end{equation}
and the peak photon energy is $\varepsilon_{\rm peak} \approx 3.92 k_{\rm B} T_{\rm c,bo} \times (R_{\rm sat}/R_{\ast})^{-1} (R_{\rm ph}/R_{\rm sat})^{-2/3}$, or
\begin{equation}\label{eq:epeak}
\varepsilon_{\rm peak} =  2.8 k_{\rm B} T_{\rm c,bo} \times (\tau_{\rm c, bo}{}^2 \eta_{\rm c})^{-1/6}.  
\end{equation}
In Table \ref{table_1}, we show $(L_{\rm peak}, \varepsilon_{\rm peak}, t_{\rm peak})$ for each progenitor.

\begin{figure}
\includegraphics[width=90mm]{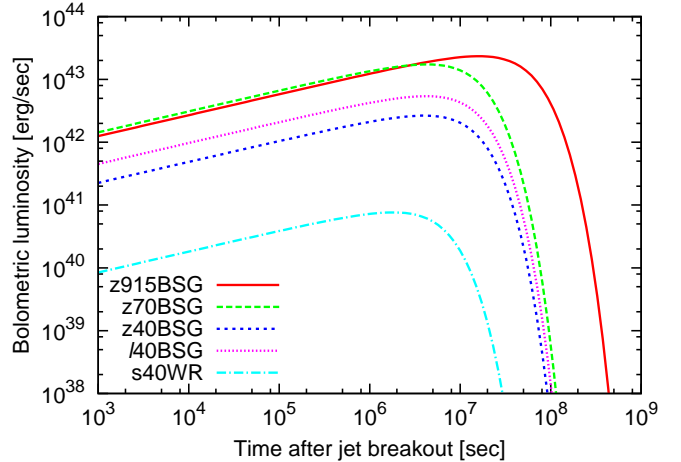}
\caption{Bolometric luminosities of CFPE from the progenitors in Table 1.  
We fix $\alpha = 1$, $\eta_{\rm j} = 6.2 \times 10^{-4}$, and $\theta_{\rm j} = 0.1$, and assume an exponential cut-off at $t = t_{\rm peak}$.}
\label{f1}
\end{figure}

Fig. \ref{f1} shows intrinsic bolometric luminosities of the cocoon-fireball photospheric emissions (CFPEs).  
We simply assume an exponential cutoff at $t = t_{\rm peak}$ for each case.  
This treatment would be justified since \cite{Peer_et_al_2006} numerically shows that the cutoff is steeper than $\propto t^{-4}$ for relativistic cocoon fireballs. 
The initial energy stored in the cocoon for z70BSG is about two times larger than that for z40BSG 
while the initial optical depth and the baryon load of z70BSG are smaller than that of z40BSG (see Table \ref{table_1}).  
Then, the cocoon of the z70BSG case becomes transparent faster than that of the z40BSG case so that the former case is brighter.  
The initial optical depth for z915BSG is $\sim$ 30 times larger than that for z70BSG while the baryon load $\eta_{\rm c}$ are similar.  
Both the initial cocoon energy and the baryon mass of the z915BSG case are larger than those of the z70BSG case.  
As a result, the evolution of the fireball looks similar, but the duration of the latter case becomes shorter.

Comparing the progenitors with different metallicity, the CFPE from the metal-poor BSG progenitors with $Z = 10^{-4} Z_{\odot}$ is similar to that of the Pop III progenitors, 
which can be understood from the fact that their pre-collapse stellar structures are also similar.  
In contrast, the cocoon emission from the WR progenitors are much dimer than the other progenitors.  
This is essentially because the timescale for the jet to break out the envelope is much shorter in the case of WR progenitors 
so that the energy stored in the cocoon is relatively small, about a tenth of the zBSG40 case.  
As a result, the peak luminosity is $\lesssim 100$ times dimmer.  
Therefore, the observation of the CFPE can be used as a diagnostic for successful Pop III GRBs and BSG GRBs.  
Moreover, one could constrain the progenitor mass from the observed parameters, $(L_{\rm peak}, \varepsilon_{\rm peak}, t_{\rm peak})$, in principle.

\begin{figure}
\includegraphics[width=90mm]{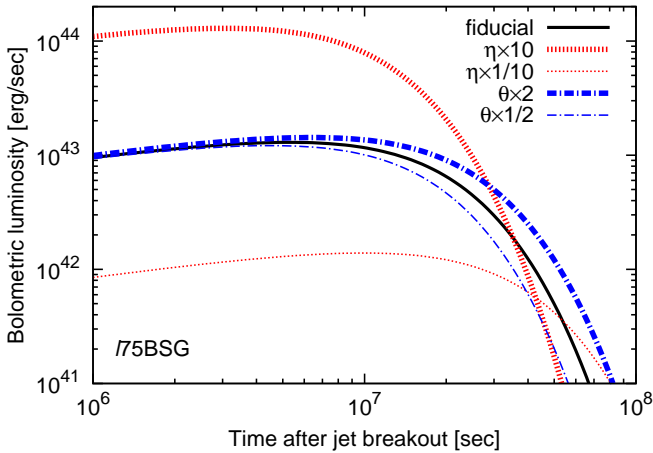}
\caption{Bolometric luminosities of CFPE from the {\it l}75BSG progenitor with different parameter set of ($\eta_{\rm j}, \theta_{\rm j}$) shown in Table 1.  
We fix $\alpha = 1$, and assume an exponential cut-off at $t = t_{\rm peak}$. }
\label{f2}
\end{figure}

Next let us mention the dependence of the CFPE on the model parameters.  
Fig. \ref{f2} shows the cocoon emission from the {\it l}75BSG progenitor with different parameter sets of $(\eta_{\rm j}, \theta)$ with fixing $\alpha = 1$ (see also Table 1).  
The solid line corresponds to our fiducial case, and the cases in which $\eta_{\rm j}$ is $10$ times larger (thick dotted line) or smaller (thin dotted line), 
and $\theta$ is $2$ times larger (thick dotted-dash line) or smaller (thin dotted-dash line) are shown.  
One can see that, by increasing $\eta_{\rm j}$, the peak luminosities increase and the durations become shorter, and by increasing $\theta$, 
the durations become longer, but the peak luminosities are almost the same.  
Using these informations, we can probe $(\eta_{\rm j},\theta)$, i.e., the efficiency and the half-opening angle of the jet inside the progenitor, 
from the observation of the cocoon emission once the progenitor is fixed, in principle.

Now we investigate the spectral evolution of CFPEs in order to discuss the detectability using current and future facilities.  
The observed fluxes in each frequency band can be described as
\begin{equation}
F_{\lambda_{\rm obs}} (t_{\rm obs}) d \lambda_{\rm obs} \approx B_{\lambda}(T_{\rm c}(t))d\lambda \cdot \frac{R(t) \Delta_R c}{\tau(t) D_{\rm L}(z){}^2}, 
\end{equation}
where $t_{\rm obs} = t(1+z)$, $\lambda_{\rm obs} = \lambda(1+z)$, and $D_{\rm L}(z)$ is the luminosity distance of the source.  
We simply assume that the spectra are black-body ones, $B_{\lambda}(T) = (2h c^2/\lambda^5) (\exp(hc/\lambda k_{\rm B}T)-1)^{-1}$ for $R_{\ast} < R < R_{\rm ph}$ 
with $h$ being the Planck constant.  
For a given wavelength, noting that the Rayleigh-Jeans law is applicable for optical and infrared radiation, $F_{\lambda_{\rm obs}} (t_{\rm obs}) \propto t_{\rm obs}{}^{7/3}$.  
In the engine rest frame, the flux at wavelength $\lambda$ takes its maximum at $h c /\lambda \approx 2.89 k_{\rm B} T_{\rm c}(t)$ as long as it occurs before $t = t_{\rm peak}$.  
This means that the observed flux takes its maximum at
\begin{eqnarray}
&{}& t_{\rm obs, peak}(\lambda_{\rm obs}) \notag \\ &{}& \approx {\rm min}\left[t_{\rm peak} (1+z), \frac{1.74 t_{\rm sat}}{(1+z)^{1/2}} \left(\frac{h c/\lambda_{\rm obs}}{k_{\rm B} T_{\rm c,bo}} \right)^{-3/2} \right], 
\end{eqnarray}
where $t_{\rm sat} = R_{\rm sat}/v_{\rm sat}$. 

\begin{figure*}[htbp]
\includegraphics[width=90mm]{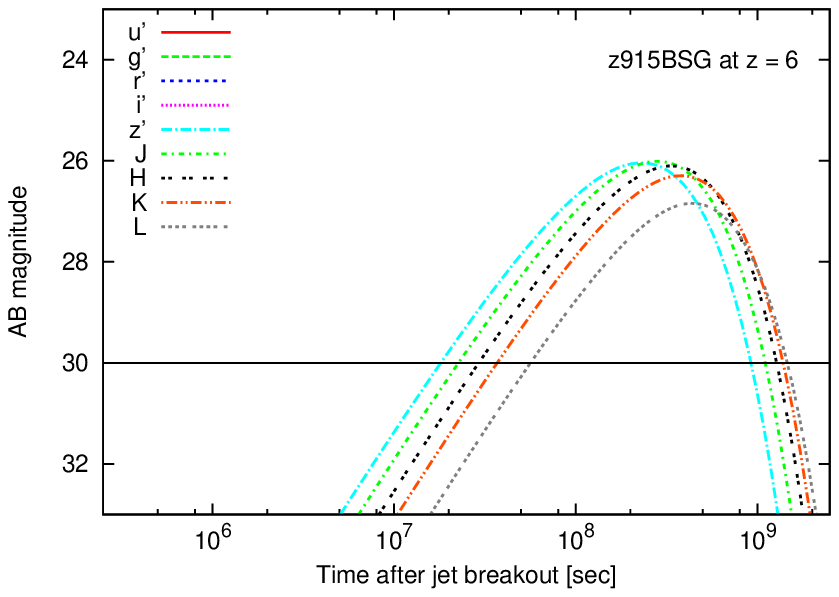}
\includegraphics[width=90mm]{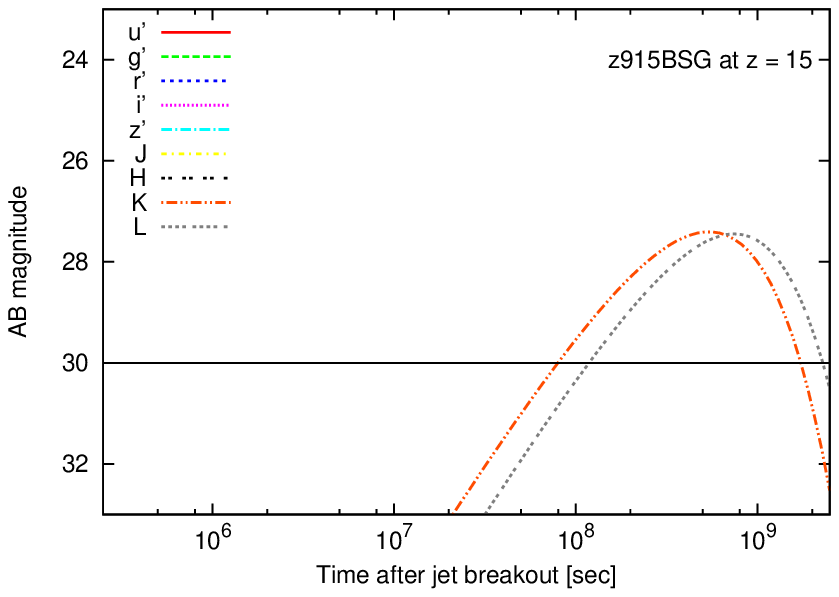}
\caption{AB magnitudes of CFPE from a
  Pop III GRB with $M_{\rm ZAMS} = 915 M_{\odot}$ at $z = 6$ (left)  and $z = 15$ (right).  We fix $\alpha = 1$, $\eta_{\rm j} = 6.2 \times 10^{-4}$, and $\theta_{\rm j} = 0.1$.  
  The horizontal line shows the anticipated $10 \sigma$-detection limit using JWST with an exposure time of $100 \mathrm{h}$.  
  The effect of the Lyman-$\alpha$ absorption is included.}
\label{f3}
\end{figure*}

\begin{figure*}[htbp]
\includegraphics[width=90mm]{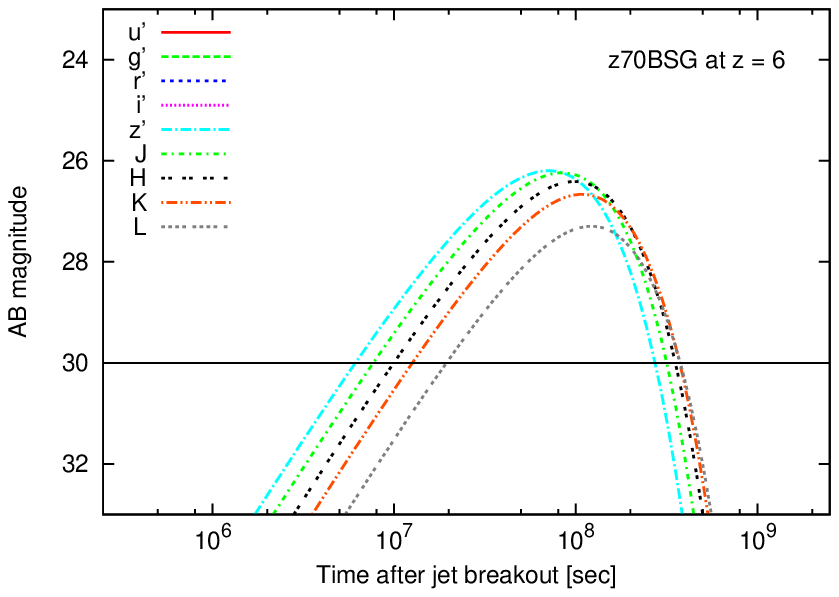}
\includegraphics[width=90mm]{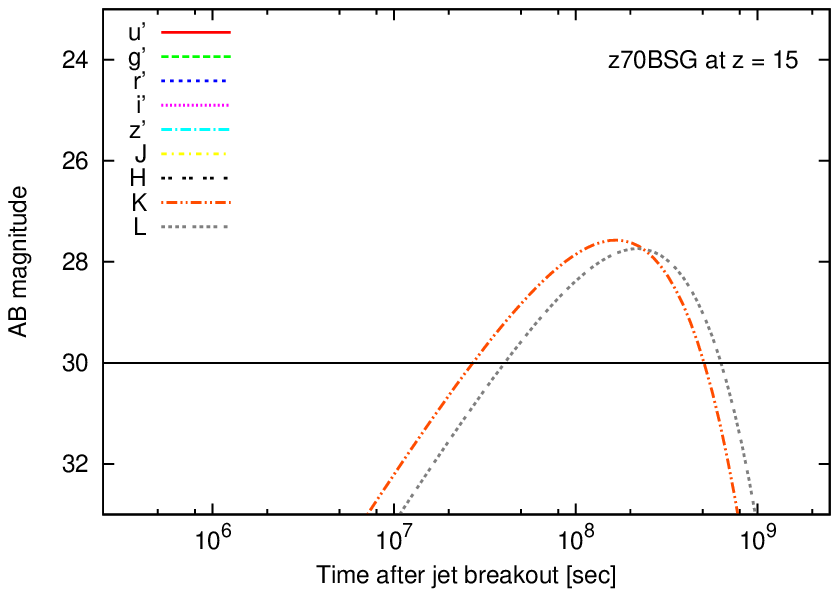}
\caption{As in Fig.\ref{f2}, but from a Pop III GRB with $M_{\rm ZAMS}  = 70 M_{\odot}$.}
\label{f4}
\end{figure*}

\begin{figure*}[htbp]
\includegraphics[width=90mm]{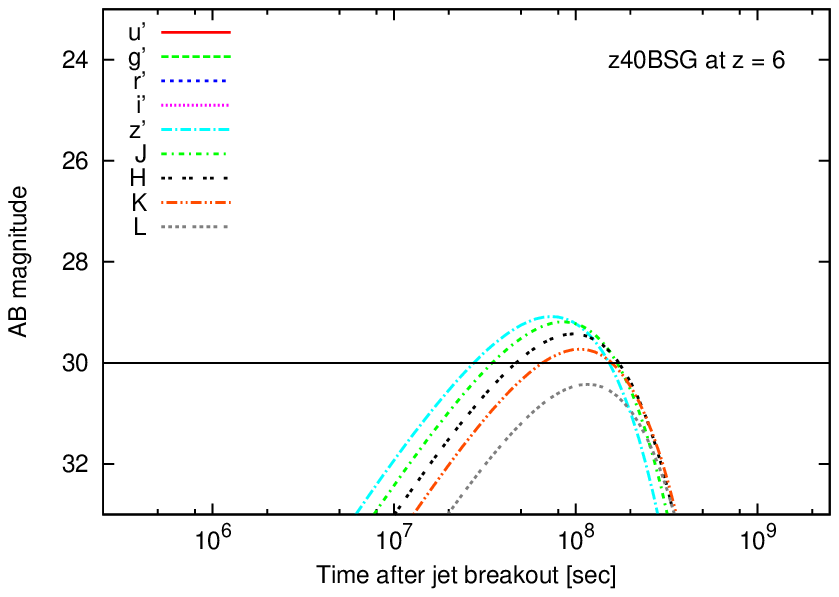}
\includegraphics[width=90mm]{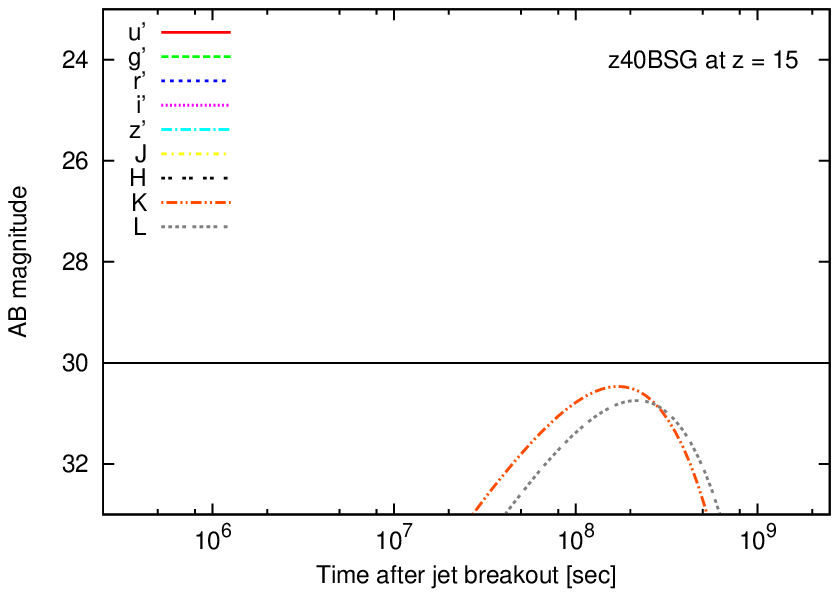}
\caption{As in Fig.\ref{f2}, but from a Pop III GRB with $M_{\rm ZAMS} = 40 M_{\odot}$.}
\label{f5}
\end{figure*}

Figs. \ref{f3}, \ref{f4} and \ref{f5} show anticipated AB magnitudes of CFPE from Pop III stars in each band of optical to infrared. 
We show the cases at $z = 6$ and $15$ including the effect of Lyman-$\alpha$ absorption.  
Photons with observed wavelength $\lambda_{\rm obs} \lesssim 0.85 {\rm \mu m}$ and $\lesssim 1.9 {\rm \mu m}$ 
are almost completely absorbed in the case of $z = 6$ and $15$, respectively, due to unionized interstellar medium.  
We neglect the extinction by intergalactic dusts \citep{Weinmann_Lilly_2005}.  
The horizontal line shows the anticipated $10 \sigma$-detection limit using JWST NIRcam with an exposure time of $100 \mathrm{h} = 3.6 \times 10^5 \ {\rm sec}$
\footnote{http://www.stsci.edu/jwst/instruments/nircam/sensitivity/table} \citep[see also][]{Zackrisson_et_al_2011}.

From Figs. \ref{f3} and \ref{f4}, one can see that JWST can detect the CFPE from Pop III GRBs with $\gtrsim 70 M_{\odot}$ at up to $z \sim 15$.  
The anticipated duration would be longer than years. 
The event rate of such Pop III GRBs is still highly uncertain.  
Based on an optimistic assumptions, the all-sky event rate could be as high as $\lesssim O(10^{5}) \ \rm yr^{-1}$ integrated over $z \gtrsim 6$ \citep{de_Souza_et_al_2011}. 
The FOV of JWST NIRcam with the shown sensitivity, $2.2 \times 2.2 \ {\rm arcmin}$ would not be wide enough for a blind search.  
Thus, a follow-up observation is needed, which is triggered by a wide-field X-ray telescope such as $\it Lobster$ and $\it EXIST$ for the prompt X ray and gamma rays
\citep{Nakauchi_et_al_2012}, or by a radio interferometers like ALMA, EVLA, LOFAR, and SKA for the radio afterglow \citep{Ioka_Meszaros_2005, Inoue_et_al_2007, Toma_et_al_2011}.

\begin{figure}
\includegraphics[width=90mm]{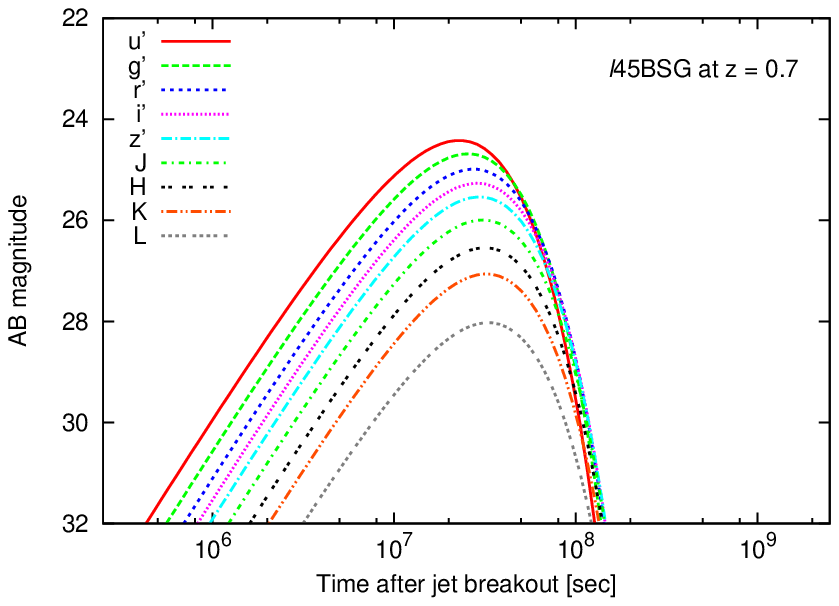}
\includegraphics[width=90mm]{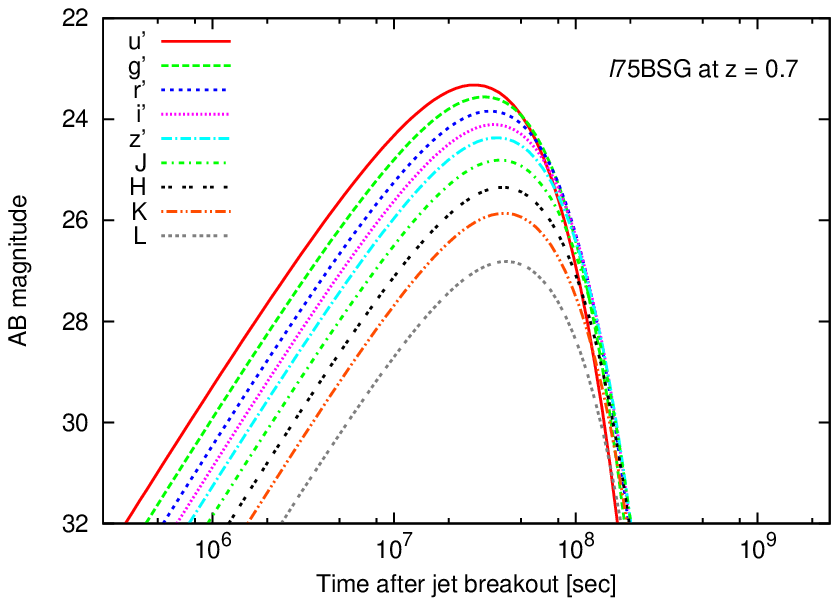}
\caption{As in Fig.\ref{f2}, but from a metal-poor BSG GRB with $40 M_{\odot}$ (top) and $75 M_{\odot}$ (bottom) at $z = 0.7$.}
\label{f6}
\end{figure}

The cocoon emissions from Pop III GRBs with $\lesssim 40 M_{\odot}$ would be relatively hard to detect at $z \gtrsim 6$ (see Fig. \ref{f5}).  
Nevertheless, such relatively low mass progenitors might exist even in a lower redshift, namely $z \lesssim 1$.
Fig. \ref{f6} shows anticipated AB magnitude from {\it l}40BSG and {\it l}75BSG at $z = 0.7$.  
In these cases, an optical/infrared counterpart from months to a year after the prompt emission would be expected.
Interestingly, an observed ultra-long GRB111209A at $z = 0.677$, which exhibited very long duration of $T_{90}\sim 2.5\times 10^{4}$ sec, 
might be originated from such a BSG progenitor in the local universe \citep{Gendre_et_al_2012}.  
Based on our scenario, it might have been recently rebrightening in optical/infrared bands up to an AB magnitude of $\lesssim 26$, which could be detectable by current facilities.

\section{Summary and discussion}\label{sec:sum_dis}
We have investigated the possibility that hot-plasma cocoons associated with successful-GRB jets produce detectable signals.  
By using a simple model, we have calculated the formation and the evolution of cocoon before and after the jet breakouts, 
and the photospheric emissions from the cocoon for Pop III or metal-poor BSG, and also WR progenitors.  
We have found that the brightness can be highly progenitor-dependent, 
i.e., the bolometric luminosity of the Pop III or BSG progenitors can be $\gtrsim 100$ times larger than that of the WR progenitor.  
Thus, the cocoon-fireball photospheric emission (CFPE) can be a diagnostic of Pop III or BSG GRBs.  
We have shown that CFPE from Pop III GRBs even at $z \sim 15$ can be detectable in infrared bands as transients of years using JWST.
We have also considered the possibility that GRBs from metal-poor BSGs occur at relatively lower redshift, $z \lesssim 1$
\footnote{Actually, \cite{Tornatore_et_al_2007} showed that Pop III star formation could last up to $z\sim 2$.}.  
In these cases, our model predicts rebrightenings in optical/infrared bands from months to a year after the GRBs.  
Based on these results, we have proposed that an observed ultra-long GRB111209A has been potentially followed by the CFPE recently up to an AB magnitude of $\lesssim 26$, 
which could be detectable. 

The local event rate of ultra-long GRB is estimated as $\sim 9 \times 10^{-3} \ \rm Gpc^{-3} yr^{-1}$ \citep{Gendre_et_al_2012} without taking into account the beaming effect.  
Since CFPE can be seen from off-axis direction of the jet, the all-sky rate of the events can be as high as $\sim 2 (\theta_{\rm j}/0.1)^{-2} \ \rm Gpc^{-3} yr^{-1}$, 
which is still much lower than a SN Ibc rate of $\sim 2 \times 10^4 \ \rm Gpc^{-3} yr^{-1}$ \citep{Madau_et_al_1998}.  
Nevertheless, they might have been already observed as a peculiar-type SNe given that 
the duration is long $(\lesssim 100 \ \rm days)$ and the total radiation energy is large $(\gtrsim 10^{50} \ \rm erg)$.  
Especially, in the cases with a relatively energetic jet where $\eta_{\rm j}$ is 10 times larger than our fiducial value, 
total radiation energy becomes as large as $\lesssim 10^{51} {\rm erg}$ (see Table 1), which can be comparable to that of the observed superluminous supernovae (SLSNe).  
The energy source of SLSNe is still highly controversial \citep[][and references therein]{GalYam_2012}.  
Given the local SLSN rate of $\sim 10 \ \rm Gpc^{-3} yr^{-1}$, some of them might be originated from cocoon-fireballs associated with metal-poor BSG GRBs.

\begin{figure}
\includegraphics[width=90mm]{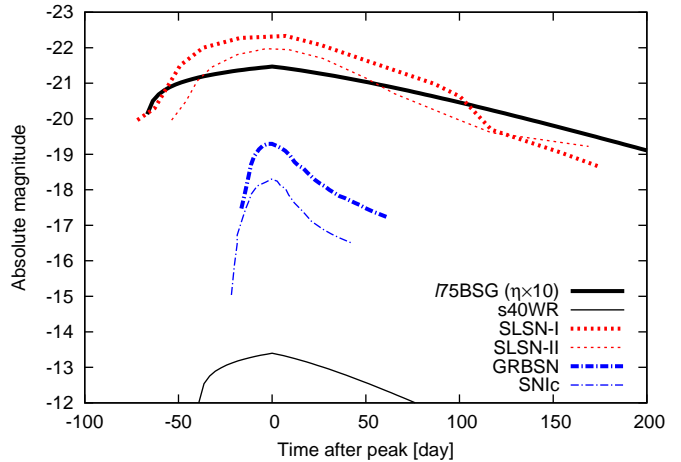}
\caption{Light curves of CFPE and SNe.  
The cocoon emission from the {\it l}75BSG progenitor with $\eta_{\rm j} = 6.2 \times 10^{-3}$ and the s40WR progenitor are compared with $\eta_{\rm j} = 6.2 \times 10^{-4}$ 
are compared with SLSN-I (PTF09cnd), SLSN-II (SN2006gy), SNIc associated with GRB (1998bw), and an average of the  observed SNIbc.  
The cocoon emissions are plotted with the bolometric magnitude, and the SNe are with the R-band magnitude.}
\label{f7}
\end{figure}

In Fig. \ref{f7}, we compare the light curve of cocoon emissions with that of various type SNe.  
We plot the bolometric magnitude of the cocoon emissions; the {\it l}75BSG model with $10$ times larger than our fiducial value for $\eta_{\rm j}$ (thick solid line) 
and the s40WR model (thin solid line).  
For the SNe, we plot the absolute magnitude in the observed R band.  
We show a class I SLSN \citep[PTF09cnd; thick dotted line;][]{Quimby_et_al_2011}, a class II SLSN \citep[SN2006gy; thin dotted line;][]{Smith_et_al_2007}, 
a type Ic SNe associated with GRB or hypernova \citep[SN1998bw; thick dotted-dash line;][]{Galama_et_al_1998}, 
and an average of the observed type Ibc SNe \citep[thin dotted-dash line;][]{Drout_et_al_2011}.

From Fig. \ref{f7}, one can find that the CFPE from metal-poor BSG GRB can be similar to SLSNe at least in terms of the energetics and the timescale.  
In order to distinguish between, or identify the cocoon emission and such a class of SNe, more detailed radiation transfer calculations of the cocoon emission is required, 
which is beyond the scope of this paper.  
Fig. \ref{f7} also shows explicitly that the cocoon emission from the WR progenitor is much dimmer than the observed GRB SNe, or a typical type Ibc SNe.
We can expect that the cocoon emissions associated with WR GRBs are hard to detect since they would be hidden by possibly associated SNe or the host galaxy.

Cocoon emissions are observationally characterized by three parameters $(L_{\rm peak}, \varepsilon_{\rm peak}, t_{\rm peak})$, 
from which one can obtain constraints between four physical parameters of the cocoon at the jet breakout $(R_\ast, E_{\rm c,bo}, M_{\rm c,bo}, V_{\rm c,bo})$.  
In our prescription, the latter three can be derived from the phenomenological parameters $(\alpha, \eta_{\rm j}, \theta_{\rm j})$.  
For a fixed parameter set of $(\alpha, \eta_{\rm j}, \theta_{\rm j})$, the observed characteristics highly depends on the progenitor mass, as shown in Fig. \ref{f1}. 
This means that one could constrain the mass of the Pop III or BSG progenitor by detecting the CFPE.  
On the other hand, if the progenitor mass is fixed, one can constrain $(\eta_{\rm j}, \theta_{\rm j})$ from the observed light curve (see Fig.\ref{f2}).  
This means that one can potentially probe, e.g., the accretion disk formation, the jet production and propagation {\it before} the jet breakout, 
which otherwise cannot be seen only using electromagnetic signals of the prompt emissions.  
The observation of prompt gamma-ray and afterglow emissions from the relativistic jet could give 
a stronger constraint on $(\eta_{\rm j},\theta_{\rm j})$ and also $\alpha$ combined with the detection (or even non-detection) of cocoon emissions.

Although our calculations are based on several simplified assumptions, we can expect the results are qualitatively correct.  
Our model parameters $(\alpha, \eta_{\rm j}, \theta_{\rm j})$, which are generally progenitor- and time-dependent, 
could be refined by numerical studies on the accretion-disk formation and the jet propagation in the collapsing stars.  

In this paper, we assume that the inner and outer part of cocoon, i.e., shocked stellar matter and shocked jet matter,
are fully mixed (Eq. (\ref{eq:Mc})).  This estimate gives a maximum baryon loading on the cocoons.  
Let us consider the cases where the mixing of the cocoon is suppressed by a factor $\xi < 1$, i.e., $M_{\rm c,bo} \rightarrow \xi M_{\rm c,bo}$. 
The baryon loading and the optical depth of the cocoon at the jet breakout scales as $\eta_{\rm c} \propto \xi^{-1}$ and $\tau_{\rm c,bo} \propto \xi$, respectively. 
Then, from Eq. (\ref{eq:tpeak}), (\ref{eq:Lpeak}), and (\ref{eq:epeak}), the CFPE becomes brighter with a higher peak energy and a shorter duration as 
$t_{\rm peak} \sim \xi^{3/4}$, $L_{\rm peak} \sim \xi^{-11/12}$, and $\varepsilon_{\rm peak} \sim \xi^{-1/6}$.
For example, in the case of the s40WR progenitor with $\xi = 0.01$, the CFPE is characterized by 
$t_{\rm peak} \sim 7.9 \times 10^4 \ \rm sec$, $L_{\rm peak} \sim 4.6 \times 10^{42} \ \rm erg/sec$, and $\varepsilon_{\rm peak} \sim 48 \ \rm eV$. 
Such emissions from mildly-relativistic cocoon fireball are discussed in e.g., \cite{Peer_et_al_2006}.

The dynamics of cocoons just after the jet breakout, i.e., $R_{\ast} \leq R \leq R_{\rm sat}$, would be more complex than our model and can also be refined with the aid of numerical simulations.  
As for the coasting phase of the cocoon fireball, i.e., $R_{\rm sat} \leq R \leq R_{\rm ph}$, more detailed radiation transfer calculations 
are needed to distinguish cocoon emissions from other competing sources e.g., various types of SNe and galaxies.

\acknowledgments

KK thanks P\'eter M\'esz\'aros, Kohta Murase, BinBing Zhang, and P\'eter Veres for valuable discussions.  
This work is supported in part by a JSPS fellowship for research abroad, NASA NNX09AL40G, 
the Grant-in-Aid from the Ministry of Education, Culture, Sports, Science and Technology (MEXT) of Japan, 
No.23540305 (TN), No.24103006 (TN), No.23840023 (YS), MEXT HPCI STRATEGIC PROGRAM, 
and by the Grant-in-Aid for the global COE program {\it The Next Generation of Physics, Spun from Universality and Emergence} at Kyoto University.

After the submission of this paper, \cite{Levan_et_al_2013} have reported 
that GRB111209A was followed by a bright optical/IR emission of an AB magnitude of $\lesssim 24$ from $10$ to $100$ days after the prompt emission. 
Although the observed spectrum is softer than the prediction of our model here, 
the luminosity and the duration are roughly consistent with the CFPE from a metal-poor BSG progenitor with a relatively energetic jet ($\eta_{\rm j} \times 10$).


\begin{thebibliography}{60}

\bibitem[{{Blandford} \& {Znajek}(1977)}]{Blandford_Znajek_1977}
{Blandford}, R.~D., \& {Znajek}, R.~L. 1977, \mnras, 179, 433

\bibitem[{{Bromberg} {et~al.}(2011){Bromberg}, {Nakar}, {Piran}, \&
  {Sari}}]{Bromberg_et_al_2011}
{Bromberg}, O., {Nakar}, E., {Piran}, T., \& {Sari}, R. 2011, \apj, 740, 100

\bibitem[{{Bromm} {et~al.}(1999){Bromm}, {Coppi}, \&
  {Larson}}]{Bromm_et_al_1999}
{Bromm}, V., {Coppi}, P.~S., \& {Larson}, R.~B. 1999, \apjl, 527, L5

\bibitem[{{Chen} \& {Beloborodov}(2007)}]{Chen_Beloborodov_2007}
{Chen}, W.-X., \& {Beloborodov}, A.~M. 2007, \apj, 657, 383

\bibitem[{{Clark} {et~al.}(2011){Clark}, {Glover}, {Klessen}, \&
  {Bromm}}]{Clark_et_al_2011}
{Clark}, P.~C., {Glover}, S.~C.~O., {Klessen}, R.~S., \& {Bromm}, V. 2011,
  \apj, 727, 110

\bibitem[{{de Souza} {et~al.}(2011){de Souza}, {Yoshida}, \&
  {Ioka}}]{de_Souza_et_al_2011}
{de Souza}, R.~S., {Yoshida}, N., \& {Ioka}, K. 2011, \aap, 533, A32

\bibitem[{{Drout} {et~al.}(2011){Drout}, {Soderberg}, {Gal-Yam}, {Cenko},
  {Fox}, {Leonard}, {Sand}, {Moon}, {Arcavi}, \& {Green}}]{Drout_et_al_2011}
{Drout}, M.~R., {Soderberg}, A.~M., {Gal-Yam}, A., {et~al.} 2011, \apj, 741, 97

\bibitem[{{Fryer} {et~al.}(2001){Fryer}, {Woosley}, \&
  {Heger}}]{Fryer_et_al_2001}
{Fryer}, C.~L., {Woosley}, S.~E., \& {Heger}, A. 2001, \apj, 550, 372

\bibitem[{{Gal-Yam}(2012)}]{GalYam_2012}
{Gal-Yam}, A. 2012, Science, 337, 927

\bibitem[{{Galama} {et~al.}(1998){Galama}, {Vreeswijk}, {van Paradijs},
  {Kouveliotou}, {Augusteijn}, {B{\"o}hnhardt}, {Brewer}, {Doublier},
  {Gonzalez}, {Leibundgut}, {Lidman}, {Hainaut}, {Patat}, {Heise}, {in't Zand},
  {Hurley}, {Groot}, {Strom}, {Mazzali}, {Iwamoto}, {Nomoto}, {Umeda},
  {Nakamura}, {Young}, {Suzuki}, {Shigeyama}, {Koshut}, {Kippen}, {Robinson},
  {de Wildt}, {Wijers}, {Tanvir}, {Greiner}, {Pian}, {Palazzi}, {Frontera},
  {Masetti}, {Nicastro}, {Feroci}, {Costa}, {Piro}, {Peterson}, {Tinney},
  {Boyle}, {Cannon}, {Stathakis}, {Sadler}, {Begam}, \&
  {Ianna}}]{Galama_et_al_1998}
{Galama}, T.~J., {Vreeswijk}, P.~M., {van Paradijs}, J., {et~al.} 1998, \nat,
  395, 670

\bibitem[{{Gendre} {et~al.}(2012){Gendre}, {Stratta}, {Atteia}, {Basa},
  {Bo{\"e}r}, {Coward}, {Cutini}, {D'Elia}, {Howell}, {Klotz}, \&
  {Piro}}]{Gendre_et_al_2012}
{Gendre}, B., {Stratta}, G., {Atteia}, J.~L., {et~al.} 2012, ArXiv e-prints

\bibitem[{{Heger} {et~al.}(2003){Heger}, {Fryer}, {Woosley}, {Langer}, \&
  {Hartmann}}]{Heger_et_al_2003}
{Heger}, A., {Fryer}, C.~L., {Woosley}, S.~E., {Langer}, N., \& {Hartmann},
  D.~H. 2003, \apj, 591, 288

\bibitem[{{Heger} \& {Woosley}(2010)}]{Heger_Woosley_2010}
{Heger}, A., \& {Woosley}, S.~E. 2010, \apj, 724, 341

\bibitem[{{Hosokawa} {et~al.}(2011){Hosokawa}, {Omukai}, {Yoshida}, \&
  {Yorke}}]{Hosokawa_et_al_2011}
{Hosokawa}, T., {Omukai}, K., {Yoshida}, N., \& {Yorke}, H.~W. 2011, Science,
  334, 1250

\bibitem[{{Inoue} {et~al.}(2007){Inoue}, {Omukai}, \&
  {Ciardi}}]{Inoue_et_al_2007}
{Inoue}, S., {Omukai}, K., \& {Ciardi}, B. 2007, \mnras, 380, 1715

\bibitem[{{Ioka} \& {M{\'e}sz{\'a}ros}(2005)}]{Ioka_Meszaros_2005}
{Ioka}, K., \& {M{\'e}sz{\'a}ros}, P. 2005, \apj, 619, 684

\bibitem[{{Kawanaka} {et~al.}(2012){Kawanaka}, {Piran}, \&
  {Krolik}}]{Kawanaka_et_al_2012}
{Kawanaka}, N., {Piran}, T., \& {Krolik}, J.~H. 2012, ArXiv e-prints

\bibitem[{{Komissarov} \& {Barkov}(2010)}]{Komissarov_Barkov_2010}
{Komissarov}, S.~S., \& {Barkov}, M.~V. 2010, \mnras, 402, L25

\bibitem[{{Kumar} {et~al.}(2008){Kumar}, {Narayan}, \&
  {Johnson}}]{Kumer_et_al_2008}
{Kumar}, P., {Narayan}, R., \& {Johnson}, J.~L. 2008, \mnras, 388, 1729

\bibitem[{{Lazzati} \& {Begelman}(2005)}]{Lazzati_Begelman_2005}
{Lazzati}, D., \& {Begelman}, M.~C. 2005, \apj, 629, 903

\bibitem[{{Levan} {et~al.}(2013){Levan}, {Tanvir}, {Starling}, {Wiersema},
  {Page}, {Perley}, {Schulze}, {Wynn}, {Chornock}, {Hjorth}, {Cenko},
  {Fruchter}, {O'Brien}, {Brown}, {Tunnicliffe}, {Malesani}, {Jakobsson},
  {Watson}, {Berger}, {Bersier}, {Cobb}, {Covino}, {Cucchiara}, {de Ugarte
  Postigo}, {Fox}, {Gal-Yam}, {Goldoni}, {Gorosabel}, {Kaper}, {Kruehler},
  {Karjalainen}, {Osborne}, {Pian}, {Sanchez-Ramirez}, {Schmidt}, {Skillen},
  {Tagliaferri}, {Thone}, {Vaduvescu}, {Wijers}, \&
  {Zauderer}}]{Levan_et_al_2013}
{Levan}, A.~J., {Tanvir}, N.~R., {Starling}, R.~L.~C., {et~al.} 2013, ArXiv
  e-prints

\bibitem[{{MacFadyen} \& {Woosley}(1999)}]{MacFadyen_Woosley_1999}
{MacFadyen}, A.~I., \& {Woosley}, S.~E. 1999, \apj, 524, 262

\bibitem[{{Madau} {et~al.}(1998){Madau}, {della Valle}, \&
  {Panagia}}]{Madau_et_al_1998}
{Madau}, P., {della Valle}, M., \& {Panagia}, N. 1998, \mnras, 297, L17

\bibitem[{{Matzner}(2003)}]{Matzner_2003}
{Matzner}, C.~D. 2003, \mnras, 345, 575

\bibitem[{{McKee} \& {Tan}(2008)}]{McKee_Tan_2008}
{McKee}, C.~F., \& {Tan}, J.~C. 2008, \apj, 681, 771

\bibitem[{{McKinney} \& {Gammie}(2004)}]{McKinney_Gammie_2004}
{McKinney}, J.~C., \& {Gammie}, C.~F. 2004, \apj, 611, 977

\bibitem[{{M{\'e}sz{\'a}ros}(2012)}]{Meszaros_2012}
{M{\'e}sz{\'a}ros}, P. 2012, ArXiv e-prints

\bibitem[{{M{\'e}sz{\'a}ros} \& {Rees}(2001)}]{Meszaros_Rees_2001}
{M{\'e}sz{\'a}ros}, P., \& {Rees}, M.~J. 2001, \apjl, 556, L37

\bibitem[{{M{\'e}sz{\'a}ros} \& {Rees}(2010)}]{Meszaros_Rees_2010}
---. 2010, \apj, 715, 967

\bibitem[{{M{\'e}sz{\'a}ros} \& {Waxman}(2001)}]{Meszaros_Waxman_2001}
{M{\'e}sz{\'a}ros}, P., \& {Waxman}, E. 2001, Physical Review Letters, 87,
  171102

\bibitem[{{Nagakura} {et~al.}(2012){Nagakura}, {Suwa}, \&
  {Ioka}}]{Nagakura_et_al_2011}
{Nagakura}, H., {Suwa}, Y., \& {Ioka}, K. 2012, \apj, 754, 85

\bibitem[{{Nakamura} \& {Umemura}(2002)}]{Nakamura_Umemura_2002}
{Nakamura}, F., \& {Umemura}, M. 2002, \apj, 569, 549

\bibitem[{{Nakauchi} {et~al.}(2012){Nakauchi}, {Suwa}, {Sakamoto}, {Kashiyama},
  \& {Nakamura}}]{Nakauchi_et_al_2012}
{Nakauchi}, D., {Suwa}, Y., {Sakamoto}, T., {Kashiyama}, K., \& {Nakamura}, T.
  2012, \apj, 759, 128

\bibitem[{{Narayan} {et~al.}(2001){Narayan}, {Piran}, \&
  {Kumar}}]{Narayan_et_al_2001}
{Narayan}, R., {Piran}, T., \& {Kumar}, P. 2001, \apj, 557, 949

\bibitem[{{Ohkubo} {et~al.}(2009){Ohkubo}, {Nomoto}, {Umeda}, {Yoshida}, \&
  {Tsuruta}}]{Ohkubo_et_al_2009}
{Ohkubo}, T., {Nomoto}, K., {Umeda}, H., {Yoshida}, N., \& {Tsuruta}, S. 2009,
  \apj, 706, 1184

\bibitem[{{Pe'er} {et~al.}(2006){Pe'er}, {M{\'e}sz{\'a}ros}, \&
  {Rees}}]{Peer_et_al_2006}
{Pe'er}, A., {M{\'e}sz{\'a}ros}, P., \& {Rees}, M.~J. 2006, \apj, 652, 482

\bibitem[{{Popham} {et~al.}(1999){Popham}, {Woosley}, \&
  {Fryer}}]{Popham_et_al_1999}
{Popham}, R., {Woosley}, S.~E., \& {Fryer}, C. 1999, \apj, 518, 356

\bibitem[{{Quimby} {et~al.}(2011){Quimby}, {Kulkarni}, {Kasliwal}, {Gal-Yam},
  {Arcavi}, {Sullivan}, {Nugent}, {Thomas}, {Howell}, {Nakar}, {Bildsten},
  {Theissen}, {Law}, {Dekany}, {Rahmer}, {Hale}, {Smith}, {Ofek}, {Zolkower},
  {Velur}, {Walters}, {Henning}, {Bui}, {McKenna}, {Poznanski}, {Cenko}, \&
  {Levitan}}]{Quimby_et_al_2011}
{Quimby}, R.~M., {Kulkarni}, S.~R., {Kasliwal}, M.~M., {et~al.} 2011, \nat,
  474, 487

\bibitem[{{Sari} \& {Piran}(1995)}]{Sari_Piran_1995}
{Sari}, R., \& {Piran}, T. 1995, \apjl, 455, L143

\bibitem[{{Smith} {et~al.}(2007){Smith}, {Li}, {Foley}, {Wheeler}, {Pooley},
  {Chornock}, {Filippenko}, {Silverman}, {Quimby}, {Bloom}, \&
  {Hansen}}]{Smith_et_al_2007}
{Smith}, N., {Li}, W., {Foley}, R.~J., {et~al.} 2007, \apj, 666, 1116

\bibitem[{{Stacy} {et~al.}(2010){Stacy}, {Greif}, \&
  {Bromm}}]{Stacy_et_al_2010}
{Stacy}, A., {Greif}, T.~H., \& {Bromm}, V. 2010, \mnras, 403, 45

\bibitem[{{Starling} {et~al.}(2012){Starling}, {Page}, {Pe'er}, {Beardmore}, \&
  {Osborne}}]{Starling_et_al_2012}
{Starling}, R.~L.~C., {Page}, K.~L., {Pe'er}, A., {Beardmore}, A.~P., \&
  {Osborne}, J.~P. 2012, ArXiv e-prints

\bibitem[{{Suwa}(2012)}]{Suwa_2012}
{Suwa}, Y. 2012, ArXiv e-prints

\bibitem[{{Suwa} \& {Ioka}(2011)}]{Suwa_Ioka_2011}
{Suwa}, Y., \& {Ioka}, K. 2011, \apj, 726, 107

\bibitem[{{Suwa} {et~al.}(2007{\natexlab{a}}){Suwa}, {Takiwaki}, {Kotake}, \&
  {Sato}}]{Suwa_et_al_2007a}
{Suwa}, Y., {Takiwaki}, T., {Kotake}, K., \& {Sato}, K. 2007{\natexlab{a}},
  \apjl, 665, L43

\bibitem[{{Suwa} {et~al.}(2007{\natexlab{b}}){Suwa}, {Takiwaki}, {Kotake}, \&
  {Sato}}]{Suwa_et_al_2007b}
---. 2007{\natexlab{b}}, \pasj, 59, 771

\bibitem[{{Toma} {et~al.}(2011){Toma}, {Sakamoto}, \&
  {M{\'e}sz{\'a}ros}}]{Toma_et_al_2011}
{Toma}, K., {Sakamoto}, T., \& {M{\'e}sz{\'a}ros}, P. 2011, \apj, 731, 127

\bibitem[{{Toma} {et~al.}(2009){Toma}, {Wu}, \&
  {M{\'e}sz{\'a}ros}}]{Toma_et_al_2009}
{Toma}, K., {Wu}, X.-F., \& {M{\'e}sz{\'a}ros}, P. 2009, \apj, 707, 1404

\bibitem[{{Tominaga} {et~al.}(2007){Tominaga}, {Maeda}, {Umeda}, {Nomoto},
  {Tanaka}, {Iwamoto}, {Suzuki}, \& {Mazzali}}]{Tominaga_et_al_2007}
{Tominaga}, N., {Maeda}, K., {Umeda}, H., {et~al.} 2007, \apjl, 657, L77

\bibitem[{{Tornatore} {et~al.}(2007){Tornatore}, {Ferrara}, \&
  {Schneider}}]{Tornatore_et_al_2007}
{Tornatore}, L., {Ferrara}, A., \& {Schneider}, R. 2007, \mnras, 382, 945

\bibitem[{{Turk} {et~al.}(2009){Turk}, {Abel}, \& {O'Shea}}]{Turk_et_al_2009}
{Turk}, M.~J., {Abel}, T., \& {O'Shea}, B. 2009, Science, 325, 601

\bibitem[{{Weinmann} \& {Lilly}(2005)}]{Weinmann_Lilly_2005}
{Weinmann}, S.~M., \& {Lilly}, S.~J. 2005, \apj, 624, 526

\bibitem[{{Whalen} {et~al.}(2012{\natexlab{a}}){Whalen}, {Fryer}, {Holz},
  {Heger}, {Woosley}, {Stiavelli}, {Even}, \& {Frey}}]{Whalen_et_al_2012a}
{Whalen}, D.~J., {Fryer}, C.~L., {Holz}, D.~E., {et~al.} 2012{\natexlab{a}},
  ArXiv e-prints

\bibitem[{{Whalen} {et~al.}(2012{\natexlab{b}}){Whalen}, {Joggerst}, {Fryer},
  {Stiavelli}, {Heger}, \& {Holz}}]{Whalen_et_al_2012b}
{Whalen}, D.~J., {Joggerst}, C.~C., {Fryer}, C.~L., {et~al.}
  2012{\natexlab{b}}, ArXiv e-prints

\bibitem[{{Woosley}(1993)}]{Woosley_1993}
{Woosley}, S.~E. 1993, \apj, 405, 273

\bibitem[{{Woosley} \& {Heger}(2012)}]{Woosley_Heger_2012}
{Woosley}, S.~E., \& {Heger}, A. 2012, \apj, 752, 32

\bibitem[{{Woosley} {et~al.}(2002){Woosley}, {Heger}, \&
  {Weaver}}]{Woosley_et_al_2002}
{Woosley}, S.~E., {Heger}, A., \& {Weaver}, T.~A. 2002, Reviews of Modern
  Physics, 74, 1015

\bibitem[{{Yoshida} {et~al.}(2007){Yoshida}, {Oh}, {Kitayama}, \&
  {Hernquist}}]{Yoshida_et_al_2007}
{Yoshida}, N., {Oh}, S.~P., {Kitayama}, T., \& {Hernquist}, L. 2007, \apj, 663,
  687

\bibitem[{{Zackrisson} {et~al.}(2011){Zackrisson}, {Rydberg}, {Schaerer},
  {{\"O}stlin}, \& {Tuli}}]{Zackrisson_et_al_2011}
{Zackrisson}, E., {Rydberg}, C.-E., {Schaerer}, D., {{\"O}stlin}, G., \&
  {Tuli}, M. 2011, \apj, 740, 13

\bibitem[{{Zalamea} \& {Beloborodov}(2011)}]{Zalamea_Beloborodov_2011}
{Zalamea}, I., \& {Beloborodov}, A.~M. 2011, \mnras, 410, 2302

\end{thebibliography}

\end{document}